# DairySocialTwin: Auditing the video-to-graph evidence chain in cattle social networks

Yiwen Huang[1] and Suresh Raja Neethirajan[1,2,*]
[1]Faculty of Computer Science, Dalhousie University, Halifax, Nova Scotia, Canada
[2]Department of Animal Science and Aquaculture, Faculty of Agriculture, Dalhousie University, Truro, Nova Scotia, Canada
*Correspondence: Sureshraja@dal.ca

**In brief**

Huang and Raja present DairySocialTwin, a graph-centered audit framework for multi-agent video. Across three commercial dairy farms, it follows uncertainty from curated clips and inherited tracks through interaction evidence, camera-sequence identity reconciliation, observation opportunity, relational summaries, and identity-level return to source video.

**Summary**

Video-derived social graphs can appear precise while inheriting uncertainty from detection, tracking, classification, and aggregation. We present DairySocialTwin, a graph-centered audit framework applied to 203 videos (122 h 23 min) from 15 views at three commercial dairy farms. Curation retained 447 of 839 candidate clips; 400 entered a frozen source-clip index, including 71 legacy clips with unavailable farm and time metadata. Five models were compared on a development split, with an NRI-style model retained as an exploratory trade-off between class balance and dyadic output. In a 90-min manual audit, reconciliation improved switches, fragmentation, and continuous-track duration across three gap definitions, but constrained identity capacity and inherited detector errors limited claims of identity accuracy. Matched manual and reconciled summaries showed coarse agreement and assignment sensitivity. Camera-specific semantics, observation opportunity, and identity-summary-to-video trace-back made these dependencies inspectable. DairySocialTwin supports auditable relational inference without claiming permanent identity, calibrated behavior, or biological diagnosis.

**The Bigger Picture**

A social graph inferred from video compresses a long and uncertain evidence chain into a deceptively simple picture. Each node depends on detection and identity assignment; each edge depends on behavioral scoring, visibility, and the rule used to aggregate observations. DairySocialTwin makes these dependencies explicit. It preserves camera-sequence identities, invalid-object status, semantic eligibility, observation opportunity, and an identity-level route back to video context. Commercial dairy barns provide a demanding test because crowding, occlusion, viewpoint variation, and repetitive appearance occur together. The contribution is broader than livestock monitoring: it is a claim-matched audit architecture for any multi-agent system in which relational conclusions are built from imperfect observations. Here, "Twin" denotes a traceable computational representation of camera-local relational state, not a live farm-wide predictive digital twin.

## Introduction

Video-derived social graphs can appear more certain than the evidence from which they are built. Each node appears to represent an animal and each edge a relationship, yet both emerge from fallible detection, tracking, interaction inference, and aggregation. Markerless systems now estimate pose, maintain identities, and recognize behavior in crowded and socially complex scenes.[1–5] Strong performance at any one stage, however, does not establish the validity of the network. An identity switch transfers evidence between animals, track fragmentation represents one animal as several nodes, and a false merge can create an artificial hub. Aggregation then conceals these errors inside an orderly graph, where they may be mistaken for biological structure.

Commercial dairy barns make this problem acute. Camera views vary in angle, illumination, lens distortion, visible floor area, occlusion, and animal density. Coat appearance changes with posture, soiling, lighting, and partial visibility. Cattle-focused studies have advanced temporal pose estimation, whole-barn tracking, long-term recognition, open-range tracking, and cross-camera re-identification.[6–11] Recent vision systems also classify social encounters in dairy settings.[12,13] Most evaluations terminate at detections, identity trajectories, or short clips. Relational inference magnifies the consequence of every upstream error because a switch no longer represents only a tracking penalty; it reallocates an animal's observations, partners, and apparent network position.

Even perfect identity continuity cannot make an edge biologically context-free. Cattle social networks vary with health challenge, familiarity, age, parity, relatedness, and management.[14–20] Across animal-network research, sampling frequency, proximity definitions, observation effort, spatial overlap, and censoring can alter apparent density, centrality, and partner ranking.[21–26] Foundational guidance therefore treats association rules, network measures, and sampling design as components of the estimand rather than neutral preprocessing.[27,28] For video-derived networks, visibility defines opportunity: a dyad seen longer can accumulate more evidence without stronger social propensity. Meaningful comparison requires both interaction evidence and the time during which both animals were observable and geometrically eligible.

Deployment adds a second layer of uncertainty. Farms differ simultaneously in layout, camera hardware, stocking patterns, management routines, and animal appearance. Domain-generalization work shows why randomly mixed partitions may overstate performance in distinct settings,[29] while multicenter graph analysis illustrates the value of separated-site evaluation.[30] More broadly, related observations crossing data partitions or target-domain information guiding model selection can produce illusory generalizability.[31–35] Reproducible machine learning therefore requires a visible connection among the sample ledger, split manifest, configuration, predictions, and final figure,[36] supported by FAIR software and model artifacts.[37,38] Calibration also matters when scores accumulate as edge weights: an uncalibrated score may rank evidence without measuring event probability.[39,40]

Existing systems commonly optimize individual modules. DairySocialTwin instead treats the video-to-graph evidence chain as the object requiring audit. We hold inherited detection and pose outputs fixed, then ask how clip curation, interaction inference, camera-sequence identity reconciliation, semantic eligibility, observation opportunity, and aggregation shape the relational object presented for interpretation. This design isolates an underexamined transition: the conversion of uncertain video events into graph-level claims.

We distinguish four levels of relational evidence. Representational readability asks whether fragmentation has been reduced enough for inspection. Support fidelity asks whether the same nodes and nonzero edges are present. Weighted fidelity asks whether edge magnitudes, node strengths, and partner rankings agree. Mesoscopic fidelity asks whether communities and time-varying roles are preserved. The present study quantifies the first level through identity-continuity measures and examines downstream sensitivity qualitatively under matched manual and reconciled assignments. It does not claim aligned support, weighted, or mesoscopic fidelity because a complete cow-level crosswalk and aligned source adjacency arrays were unavailable.

The study therefore proceeds through claim-matched evidence tiers. First, five interaction models are compared on a frozen development index. Second, identity-continuity measures compare inherited and reconciled assignments against a manually corrected camera-sequence reference. Third, camera-local graph summaries preserve semantic eligibility and observation opportunity. Fourth, matched manual and reconciled outputs expose downstream sensitivity without treating visual resemblance as numerical graph agreement. Finally, selected identity-level summaries return to appearance times and source frames. Macro-F1 does not measure identity continuity, switch rate does not estimate adjacency error, and a similar marginal distribution does not establish partner-level agreement. Keeping those statements separate is the central discipline of the framework.

We ask which interaction model provides the most useful development-stage trade-off for graph construction, how much reconciliation improves continuity across alternative gap definitions, which downstream summaries remain sensitive to identity assignment or inherited detector contamination, and how provenance can make those dependencies inspectable. The contribution is not a new detector, an externally validated classifier, permanent animal identity, or biological diagnosis. It is a graph-centered audit architecture for determining which relational claims remain defensible as uncertainty propagates from video to network. Figure 1 summarizes this audit chain and the heterogeneous commercial-farm settings in which it was applied.

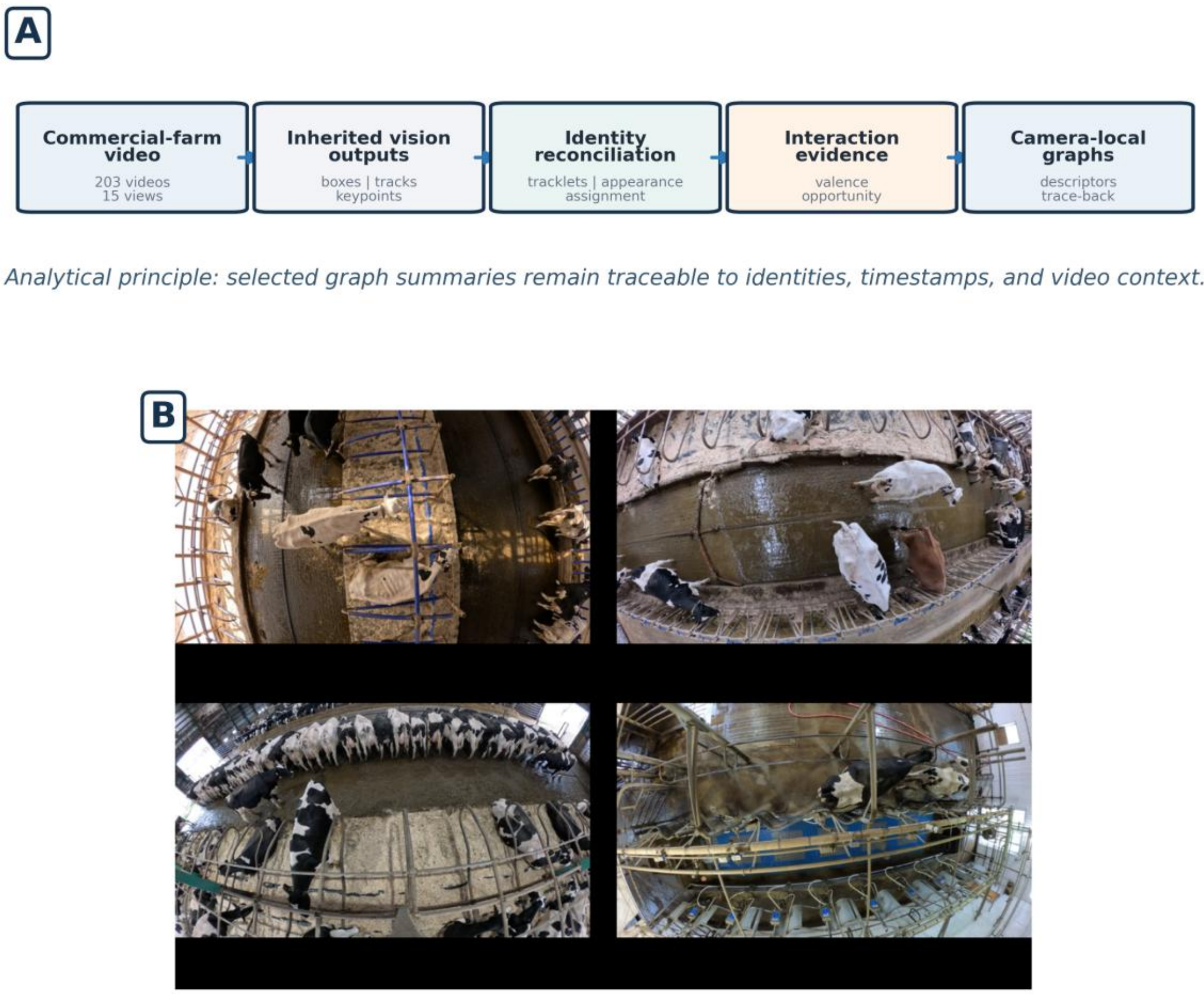

**Figure 1. Study design and commercial-farm context.** (A) The audit chain separates inherited vision outputs from camera-sequence identity reconciliation, interaction evidence, observation opportunity, graph summaries, and identity-level video trace-back. (B) Representative camera views illustrate differences in geometry, density, occlusion, and visible functional regions. Semantic maps are camera-specific and nonmetric because camera intrinsics and surveyed landmarks were unavailable.

## Results

### A frozen source-clip index separated curation from metadata availability

The study corpus comprised 203 videos totaling 122 h 23 min and 2.05 TB. Five fixed cameras at each of three commercial dairy farms provided 15 views across feeding, watering, resting, and circulation areas. Project records encompassed more than 300 cattle, but the computational identities analyzed here were deliberately limited to individual camera sequences and sessions. The herd-wide total therefore describes study coverage, not the number of animals uniquely re-identified across cameras or farms.

Candidate generation produced 839 clips. Manual review removed 264 clips without usable cattle or motion, 116 during quality control, and 12 during trajectory stabilization, leaving 447 retained archive clips. A further 47 retained clips were excluded when the experimental index was frozen, producing 400 source clips for model development. The frozen index contained 57 friendly, 107 unfriendly, and 236 no-interaction clips. Of these, 330 source clips were allocated to training and 70 to development-validation (Table 1).

Seventy-one legacy clips were included within the frozen set, not added to it. They retained interaction labels but lacked reliable farm, camera, and original TCD/tmcd timecode metadata. These clips comprised 20 friendly and 51 unfriendly examples. They were kept for overall model development because their labels and trajectories survived quality control, but they were coded as metadata unavailable and were not used to support farm- or time-stratified performance claims. This distinction corrected the earlier tendency to treat missing metadata as if it denoted a valid farm or time group.

**Table 1. Corpus and analytical-unit ledger.** The retained archive and frozen experimental index are different layers. The 75 validation segments were generated by stabilization of 70 validation source clips; they are the analytical units in Figure 2 and Table 2.

| Analytical layer | Friendly | Unfriendly | No interaction | Total | Interpretation |
|---|---|---|---|---|---|
| Candidate clips | n/a | n/a | n/a | 839 | Before manual curation |
| Retained archive clips | 66 | 145 | 236 | 447 | After curation and stabilization |
| Retained but outside frozen index | 9 | 38 | 0 | 47 | Not used in model comparison |
| Frozen experimental source clips | 57 | 107 | 236 | 400 | Training plus development-validation |
| Training source clips | 44 | 85 | 201 | 330 | Model fitting |
| Development-validation source clips | 13 | 22 | 35 | 70 | Source-clip split |
| Stabilized validation segments | 14 | 26 | 35 | 75 | Units used for confusion matrices |
| Legacy clips with unavailable metadata | 20 | 51 | 0 | 71 | Subset of the frozen 400 clips |

### Development-stage comparison revealed a transparent selection trade-off

Five trajectory-based alternatives were compared on the 75 stabilized validation segments generated from the 70 validation source clips: a direct three-class classifier, a cascaded interaction gate followed by valence classification, a non-graph Transformer, a temporal graph neural network, and an NRI-style latent-edge model. The same reference counts were used for every confusion matrix: 35 no-interaction, 14 friendly, and 26 unfriendly segments (Figure 2).

The cascaded model produced the strongest aggregate development metrics, with 55 of 75 segments correct (accuracy 0.733), macro-F1 0.697, and weighted-F1 0.729 (Table 2). Its advantage was concentrated in the unfriendly class, for which recall reached 0.885, while friendly recall was 0.429. The NRI-style model correctly classified 54 of 75 segments (accuracy 0.720), with macro-F1 0.687 and balanced accuracy 0.696. It recovered 9 of 14 friendly segments (recall 0.643), compared with 6 of 14 for the cascaded model, and exposed dyadic latent-edge outputs suited to downstream graph construction.

The NRI-style model was therefore retained as an exploratory engineering trade-off, not because it dominated every metric. The choice favored friendly-class recovery and a dyadic representation at a small cost in aggregate performance. No untouched test set was available, and the selection criterion was not prospectively registered. All classification results should consequently be read as development-stage characterization rather than an estimate of deployment performance.

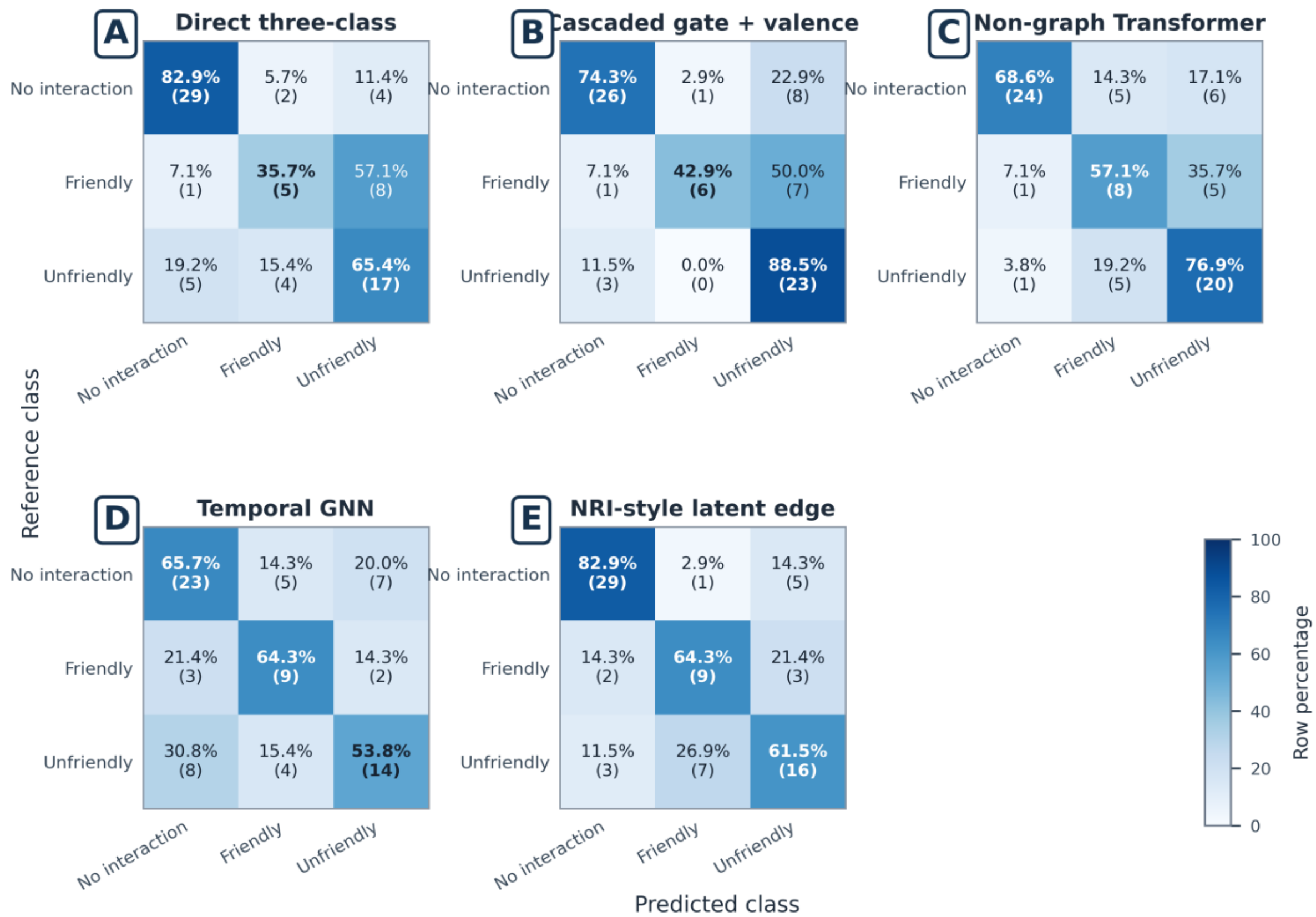

**Figure 2. Development-stage interaction-model comparison.** Row-normalized confusion matrices show percentages and integer counts for 75 stabilized validation segments derived from 70 source clips. Rows are reference classes and columns are predicted classes. The cascaded model had the strongest aggregate metrics, whereas the NRI-style model recovered more friendly segments and provided dyadic output for graph construction. No farm- or time-stratified performance result is shown because 71 frozen-index legacy clips lacked reliable grouping metadata.

**Table 2. Development-stage model performance.** Metrics were recalculated from the integer confusion matrices in Figure 2. Values describe the frozen development comparison and are not external-test estimates.

**Table 2A. Aggregate metrics.**

| Model | Accuracy | Macro-F1 | Balanced accuracy | Weighted-F1 |
|---|---|---|---|---|
| Direct three-class | 0.680 | 0.616 | 0.613 | 0.676 |
| Cascaded gate + valence | 0.733 | 0.697 | 0.685 | 0.729 |
| Non-graph Transformer | 0.693 | 0.663 | 0.675 | 0.704 |
| Temporal GNN | 0.613 | 0.600 | 0.613 | 0.614 |
| NRI-style latent edge | 0.720 | 0.687 | 0.696 | 0.723 |

**Table 2B. Class-specific F1.**

| Model | No interaction | Friendly | Unfriendly |
|---|---|---|---|
| Direct three-class | 0.829 | 0.400 | 0.618 |
| Cascaded gate + valence | 0.800 | 0.571 | 0.719 |
| Non-graph Transformer | 0.787 | 0.500 | 0.702 |
| Temporal GNN | 0.667 | 0.563 | 0.571 |
| NRI-style latent edge | 0.841 | 0.581 | 0.640 |

**Reconciliation improved continuity across three gap definitions**

Four clips from three farm-camera views were manually reviewed to construct a camera-sequence identity reference (Table 3). Together, the clips covered 2:41:12, 17,159 inherited tracklets, and 5,583 review operations. The reviewed views spanned an average of 3.54, 5.21, and 19.52 valid single-cow boxes per frame, providing progressively denser conditions. Farm 1, Camera 1 was reviewed tracklet by tracklet and supplied the quantitative continuity audit; the denser Farm 2 and Farm 3 views were reviewed in proposed groups and supported qualitative inspection.

**Table 3. Manual identity-reference inventory.** Review operations are interface decisions. For Farm 1, Camera 1, each tracklet was reviewed individually. In the denser views, proposed groups were reviewed while the actual tracklet counts were preserved.

| Farm | Camera | Clips | Duration | Review operations | Actual tracklets | Valid single-cow boxes/frame |
|---|---|---|---|---|---|---|
| 1 | 1 | 2 | 1:30:44 | 2,414 | 2,414 | 3.54 |
| 2 | 5 | 1 | 0:35:14 | 1,015 | 4,121 | 5.21 |
| 3 | 1 | 1 | 0:35:14 | 2,154 | 10,624 | 19.52 |
| **Total** | | **4** | **2:41:12** | **5,583** | **17,159** | |

Continuity was evaluated under gap thresholds of 1, 15, and 30 frames (Table 4; Figure 3A). At the strictest threshold, reconciliation reduced identity switches from 80.65 to 5.94 per hour and fragmentation from 3.19 to 0.41, reductions of 92.6% and 87.1%, respectively. Mean continuous duration increased from 9.61 to 10.19 s. At 15 frames, switch rate fell from 141.47 to 25.78 per hour, fragmentation from 5.68 to 1.14, and duration increased from 19.18 to 23.22 s. At 30 frames, switch rate fell from 159.32 to 29.08 per hour, fragmentation from 6.21 to 1.26, and duration increased from 24.98 to 33.44 s. Across the three definitions, switch reductions ranged from 81.7% to 92.6%, fragmentation reductions from 79.7% to 87.1%, and duration gains from 6.0% to 33.9%.

Stage C accepted 1,846 of 1,942 manually checked links correctly (95.06%). This result measures accepted short- and long-gap links, not errors introduced during the subsequent global assignment. Likewise, the reduction from 617 inherited identifiers to 62 reconciled outputs cannot be read as identity accuracy because 62 was the imposed Farm 1 capacity, comprising a confirmed herd-size prior of 57 plus 5 slots for residual detections. The continuity results show cleaner camera-sequence trajectories; they do not quantify false merging, identity purity, or permanent identity.

**Table 4. Gap-sensitive identity-continuity audit for Farm 1, Camera 1.** Percent changes are relative to inherited assignments. Higher continuous duration is favorable; lower switch rate and fragmentation are favorable.

**Table 4A. Identity switches per reviewed hour.**

| Gap threshold | Inherited | Reconciled | Reduction |
|---|---|---|---|
| 1 frame | 80.65 | 5.94 | 92.6% |
| 15 frames | 141.47 | 25.78 | 81.8% |
| 30 frames | 159.32 | 29.08 | 81.7% |

**Table 4B. Track fragmentation.**

| Gap threshold | Inherited | Reconciled | Reduction |
|---|---|---|---|
| 1 frame | 3.19 | 0.41 | 87.1% |
| 15 frames | 5.68 | 1.14 | 79.9% |
| 30 frames | 6.21 | 1.26 | 79.7% |

**Table 4C. Mean continuous-track duration.**

| Gap threshold | Inherited (s) | Reconciled (s) | Increase |
|---|---|---|---|
| 1 frame | 9.61 | 10.19 | 6.0% |
| 15 frames | 19.18 | 23.22 | 21.1% |
| 30 frames | 24.98 | 33.44 | 33.9% |

Appearance-similarity matrices showed coherent within-sequence blocks but also ambiguous regions consistent with repeated coat patterns and partial views (Figure 3B). Manual review further exposed upstream failure modes that identity reconciliation could not repair, including stable detections containing more than one cow. In the two Farm 1 clips, the inherited pipeline produced an average of 4.57 boxes per frame, compared with 3.54 valid single-cow boxes per frame after review. A clean identity label attached to an invalid entity remains invalid evidence.

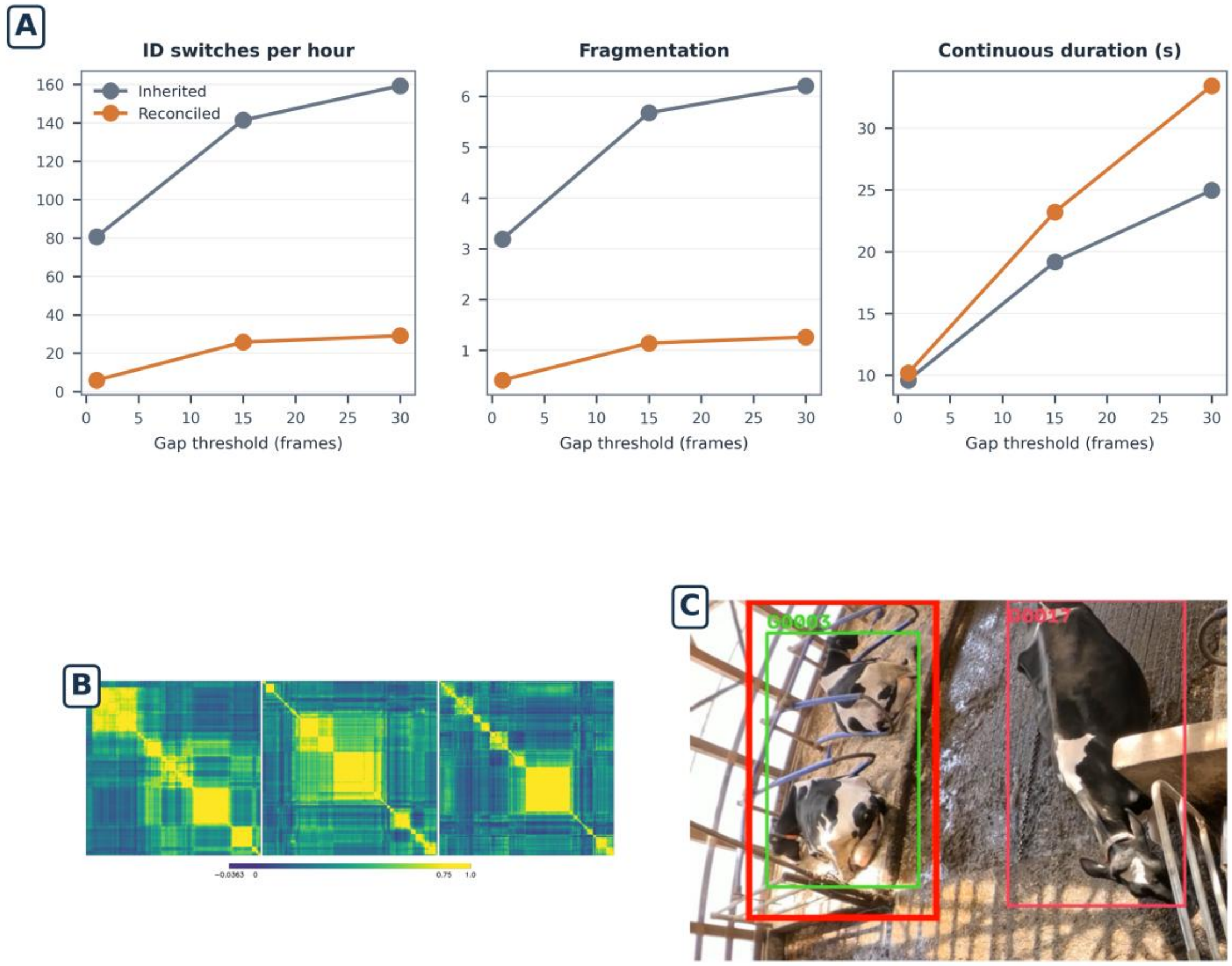

**Figure 3. Identity reconciliation and residual failure modes.** (A) Inherited and reconciled switch rates, fragmentation, and mean continuous duration under 1-, 15-, and 30-frame gap definitions in the 90-min Farm 1 audit. (B) Appearance-similarity structure used to assess within-sequence prototypes. (C) Example of an inherited detection containing multiple cows; the panel is cropped to exclude farm personnel. Reconciliation improved continuity but did not correct the semantic validity of inherited detections.

**Opportunity conditioning separated visibility from interaction evidence**

DairySocialTwin retained friendly and unfriendly score-weighted evidence as separate non-negative layers and accumulated confidence-weighted opportunity only while both identities were visible and geometrically eligible. A pair with zero opportunity was therefore marked unobserved rather than assigned a negative interaction value. Camera-specific semantic maps further excluded pairings that crossed annotated barriers and localized evidence to functional regions.

The resulting outputs provided several complementary, explicitly descriptive views (Figure 4). Region-level summaries contrasted friendly and unfriendly evidence after conditioning on opportunity (Figure 4A). A sparse adjacency panel localized relatively strong unfriendly evidence among selected anonymous identities (Figure 4B). An exploratory isolation decomposition retained camera-local alone fraction, low-friendly-strength, and partner-weight evenness rather than hiding them inside a single score (Figure 4C). Cow-level descriptors juxtaposed friendly and unfriendly strength, community co-membership stability, alone fraction, and isolation score (Figure 4D).

These panels are useful for triage and hypothesis generation, not biological diagnosis. Anonymous identifiers are local to a camera sequence, score outputs are not calibrated event probabilities, semantic zones are defined in image space, and the composite isolation index is heuristic. The plots therefore describe where model-derived evidence accumulated and which identities warrant inspection. They do not establish affinity, aggression, welfare status, or stable herd rank.

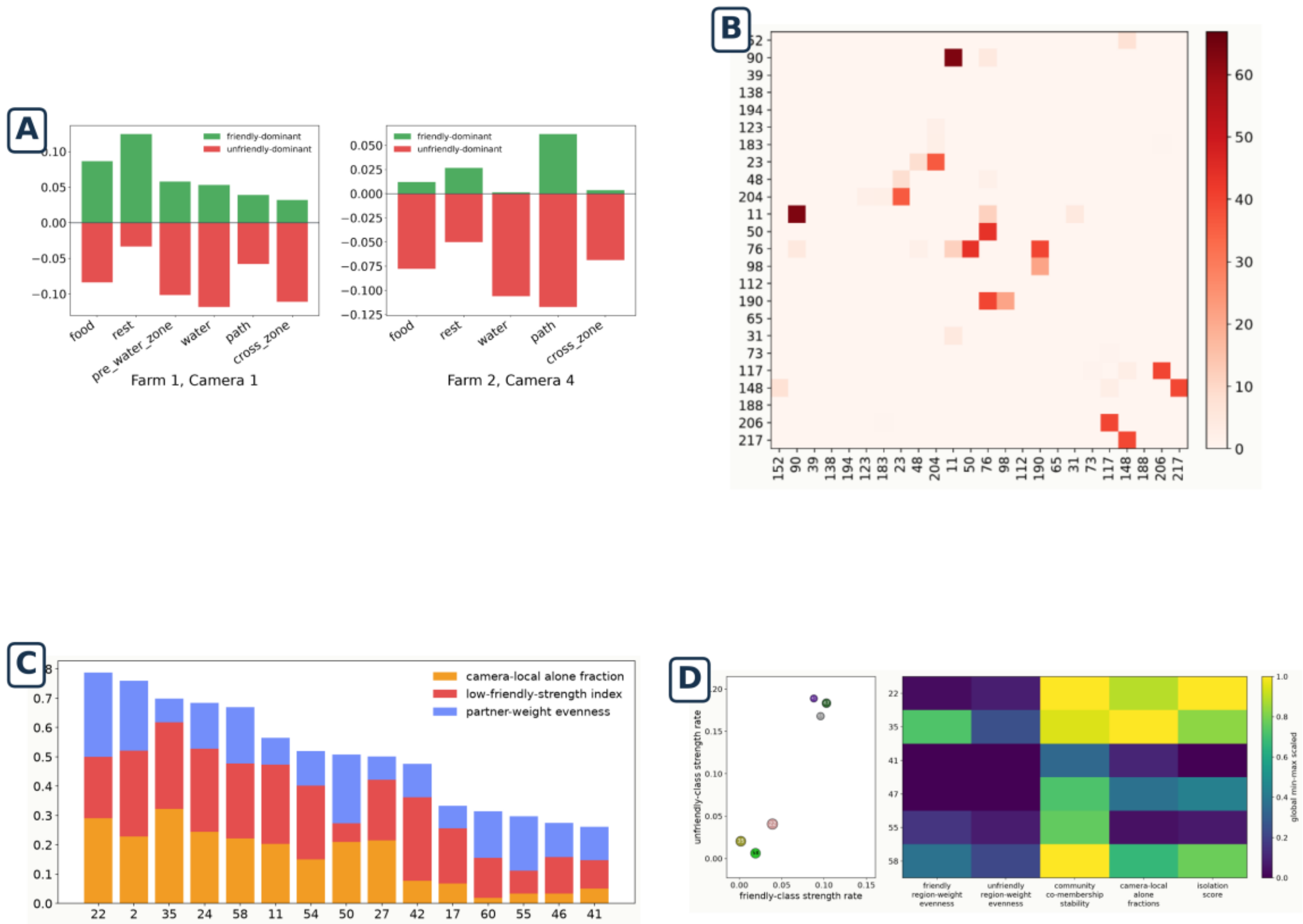

**Figure 4. Opportunity-conditioned camera-local outputs.** (A) Aggregate friendly- and unfriendly-class evidence by functional region in Farm 1, Camera 1 and Farm 2, Camera 4. (B) Unfriendly-class adjacency display for selected Farm 2, Camera 4 identities. (C) Components of the exploratory isolation index in Farm 1, Camera 1. (D) Selected identity-level descriptors in Farm 1, Camera 1. Values are model-derived, camera-local summaries and are not calibrated behavioral rates.

**Matched outputs exposed downstream sensitivity without implying graph fidelity**

For each reviewed record, downstream outputs were regenerated under manually corrected and reconciled identity assignments while the observation window, semantic map, inherited detections, and interaction scores were held fixed. The resulting trajectories often preserved the same coarse spatial envelope, and several aggregate dwell and isolation distributions were qualitatively similar. These similarities indicate representational readability: reconciliation recovered a legible summary of broad camera-local use.

The agreement did not extend uniformly to downstream relational evidence. In Farm 1, Camera

1, the relative friendly and unfriendly contributions in the pre-water region changed under manual and reconciled assignments (Figure 5B). The same region contained a high prevalence of inherited multi-cow boxes. Because only detections with usable trajectories and valid dyadic geometry contributed to interaction evidence, identity reassignment and invalid-object contamination changed which observations entered each summary. Independent identity ordering across panels further prevented row-by-row cow comparisons.

We therefore treat Figure 5 as a sensitivity analysis rather than a graph-fidelity result. A complete cow-level crosswalk and aligned source adjacency arrays were unavailable, so edge support, weighted agreement, strongest-partner retention, node-strength ranks, and community agreement were not calculated. Visual resemblance of marginal distributions cannot substitute for these quantities.

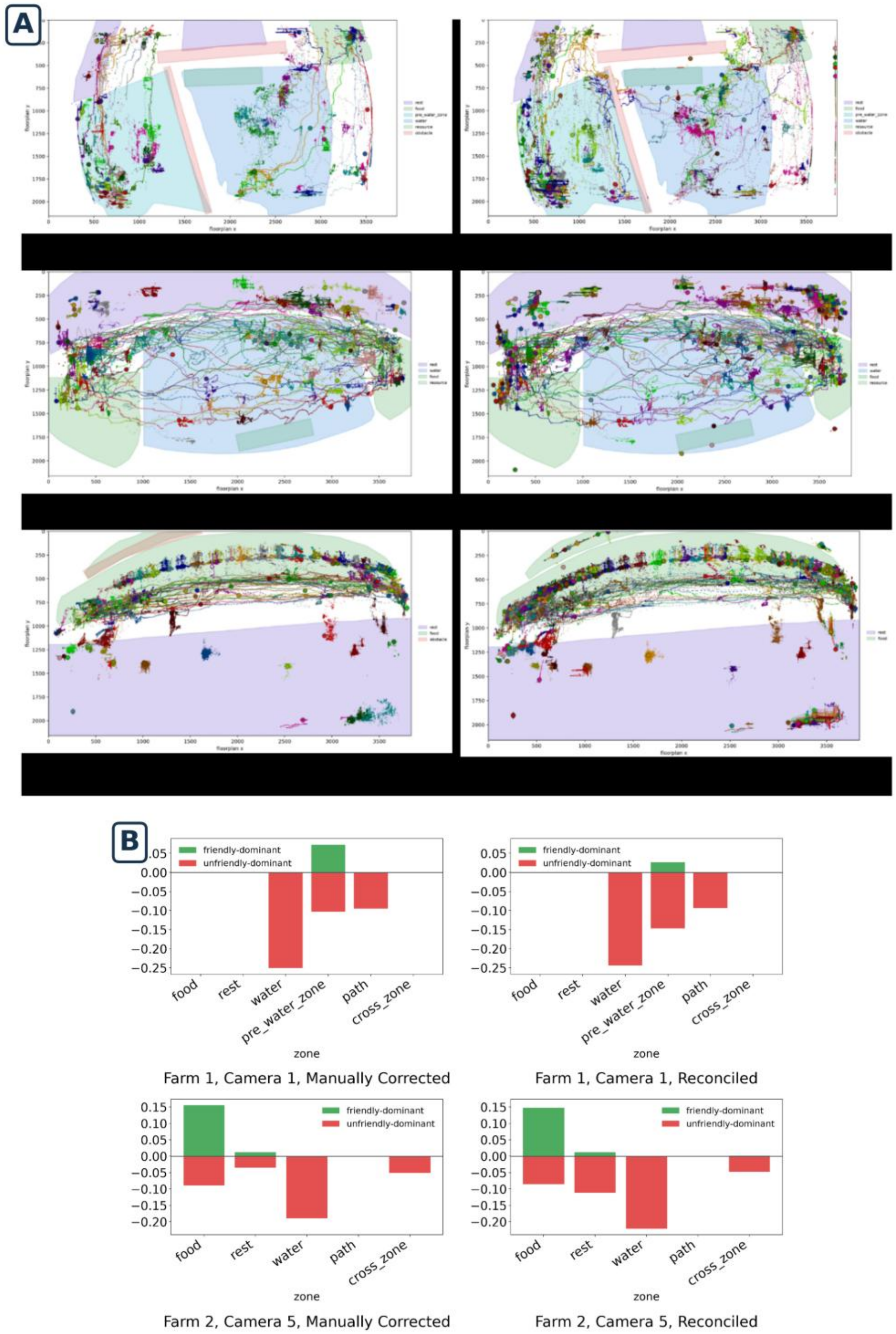

**Figure 5. Downstream sensitivity to identity assignment.** (A) Paired manually corrected and reconciled trajectory summaries across the reviewed farm-camera views. Columns show manual and reconciled assignments; rows show the three reviewed contexts. (B) Paired aggregate opportunity-conditioned regional evidence for Farm 1, Camera 1 and Farm 2, Camera 5. Panel orderings are independent and no complete cow-level crosswalk was available. The figure supports qualitative sensitivity assessment, not node- or edge-level fidelity.

**Identity-level trace-back made compressed summaries inspectable**

The review interface presented inherited occurrences and allowed a reviewer to accept an assignment, change an anonymous identity, or mark a detection invalid (Figure 6A). The analytical dashboard displayed stored camera-local summaries for selected identities (Figure 6B). Selecting an identity-level datum exposed recorded appearance intervals and the corresponding camera frame (Figure 6C).

This return path changes the role of a graph summary. An unusual point can be investigated as sustained evidence, a short high-scoring interval, a fragmented identity, or contamination from an invalid detection. The interface does not prove that the underlying prediction is correct, and it does not yet implement direct edge-to-every-contributing-frame retrieval. It provides a bounded identity-summary-to-video trace-back that makes disagreements localizable rather than hidden inside a terminal visualization.

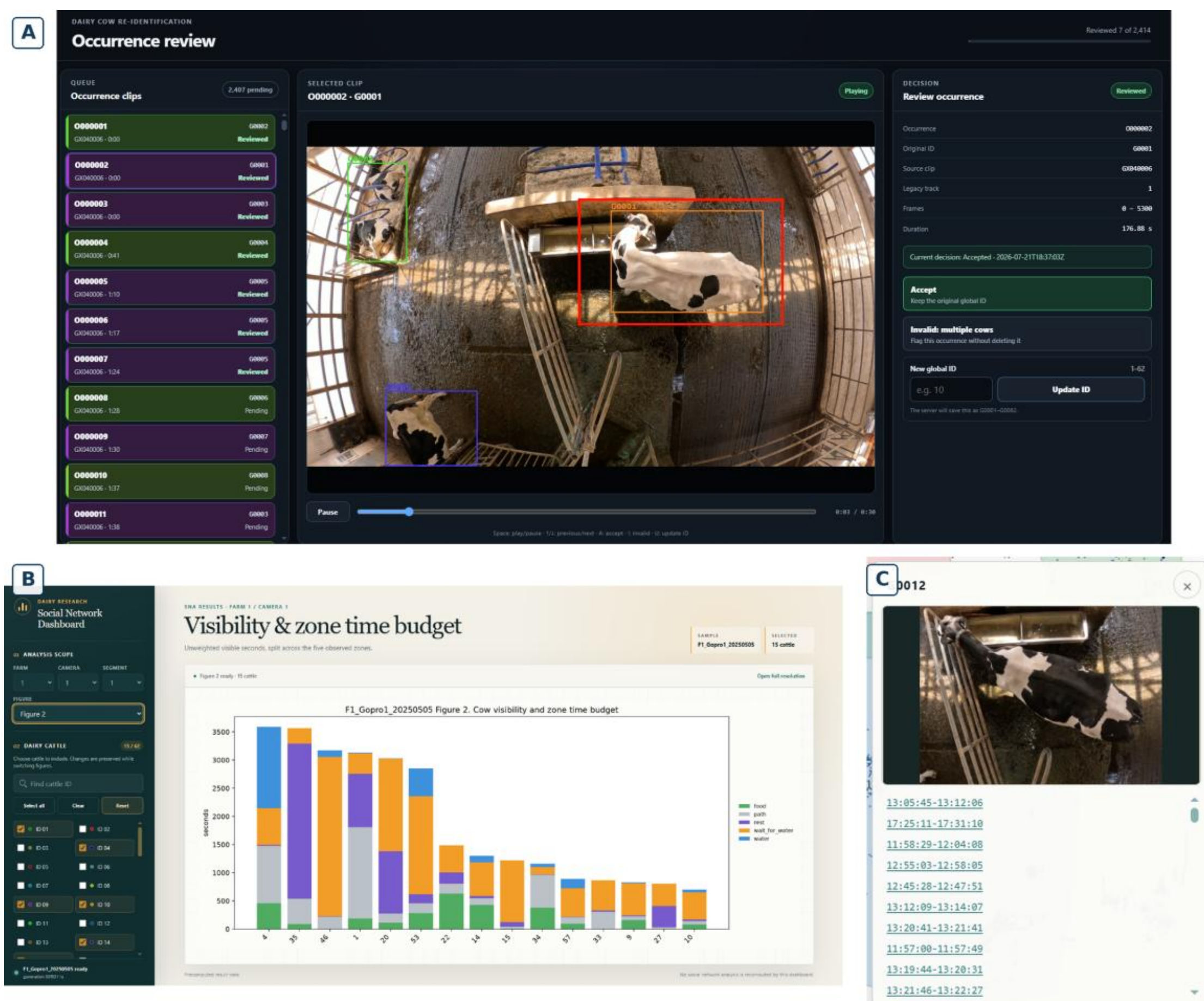

**Figure 6. Human review and evidence trace-back.** (A) Occurrence-level identity review. (B) Dashboard view of a stored camera-local summary. (C) Selected identity with appearance intervals and corresponding video context. The interface supports identity-level observational trace-back; it is not a claim of complete edge-level or computational provenance.

## Discussion

The central finding is not simply that reconciliation improves tracks. It is that track continuity and relational validity are different quantities, and the distinction becomes visible only when evaluation follows evidence to the graph. Across three gap definitions, reconciliation markedly reduced switches and fragmentation while increasing continuous-track duration. Yet manual inspection still found inherited boxes containing multiple cows and showed that region-level evidence changed when identity assignments were replaced. Cleaner trajectories made camera-local summaries legible; they did not make every node a valid cow or every edge a verified relationship.

This distinction sharpens the role of DairySocialTwin. Contemporary pose, tracking, and behavior-analysis systems provide increasingly capable components,[41–51] but their usual endpoints are coordinates, identities, or behavioral labels. A relational study asks a different question: which scientific statements survive when those outputs are aggregated? DairySocialTwin addresses that transition by retaining four dependencies that are commonly compressed away: who was assigned to an observation, whether the observation represented a valid single cow, whether the dyad was eligible to interact in the camera geometry, and how much opportunity existed to observe the pair. The result is not another terminal network visualization. It is a claim-matched audit layer between multi-agent vision and relational inference.

The development comparison establishes the first boundary. The cascaded model achieved the strongest aggregate macro-F1, whereas the NRI-style model recovered more friendly examples and generated dyadic latent-edge outputs. Selecting the latter was a transparent design trade-off, not evidence that it was universally superior. More importantly, macro-F1 near 0.69 places a clear ceiling on edge interpretation. When scores are aggregated over time, classification errors do not disappear; they are redistributed through node histories and partner sets. The framework therefore preserves friendly and unfriendly scores as model-derived evidence rather than translating them into calibrated behavioral duration.

The corrected clip ledger is itself a methodological result. Seventy-one legacy clips had usable labels but unavailable farm, camera, and original time metadata. Retaining them in the overall frozen index while excluding them from stratified claims is more informative than silently imputing a group or deleting them from the corpus total. It also prevents a common form of analytical leakage: a polished subgroup result whose grouping variable was never reliably observed. No farm- or time-specific performance claim is made here. The paper instead exposes the accounting needed to determine which comparisons the data can support.

The identity audit provides the second boundary. Reductions of 81.7% to 92.6% in switch rate and 79.7% to 87.1% in fragmentation were consistent across gap choices, which argues against a single favorable threshold. The continuous-duration gain increased as the allowable gap widened, as expected for a procedure designed to reconnect interruptions. These are substantial continuity gains against the manual camera-sequence reference. They are not equivalent to identity accuracy. Stage C link correctness does not measure false merges introduced by Stage D, and a final count of 62 identities is partly determined by the imposed capacity. Association-sensitive metrics such as HOTA, IDF1, and established multi-target benchmarks provide a broader evaluation vocabulary,[52–54] while modern association systems show the range of strategies available for recovering trajectories.[55–58] The present contribution is narrower: it demonstrates why even improved local continuity must be audited at the relational endpoint.

Observation opportunity provides the third boundary. A pair that is not jointly visible is not evidence of social avoidance, and a highly visible pair can accumulate a large raw score without unusual social propensity. Conditioning on confidence-weighted co-visibility and camera-specific eligibility separates opportunity from interaction evidence. It also preserves an explicit unobserved state when opportunity is zero. This denominator is essential in fixed-camera environments, where barriers, blind regions, crowding, and functional zones shape what can be seen. The resulting quantities remain normalized score-weighted evidence, not calibrated rates, because the underlying model probabilities were not calibrated.

Matched manual and reconciled outputs clarify the fourth boundary. Coarse trajectory envelopes and marginal distributions sometimes resembled one another, yet regional interaction proportions remained assignment-sensitive. This is precisely why visual plausibility is an insufficient validation criterion. Stronger graph evaluation requires aligned identities and numerical arrays to measure edge support, weighted errors, partner retention, node-strength ordering, and community agreement. Graph-comparison metrics and higher-order structural analyses provide suitable tools when those prerequisites are met.[59–62] In their absence, the disciplined conclusion is qualitative: reconciliation can stabilize broad camera-local summaries while leaving cow-level relational claims unresolved.

The evidence interface turns this limitation into an inspectable workflow. A selected identity-level result can be returned to appearance intervals and source frames, allowing the analyst to distinguish persistent evidence from a brief score, a fragmented assignment, or an invalid box. This is observational trace-back, not proof of correctness. Its value lies in making disagreement localizable. In a restricted-data setting, that property also allows editors, reviewers, or authorized auditors to examine sensitive evidence without requiring unrestricted publication of farm video.

The study has three principal limitations. First, interaction classification was a development-stage comparison: no untouched external test set, repeated-seed uncertainty, parent-event grouping audit, or independent annotation-agreement estimate was available. Seventy-one legacy clips lacked grouping metadata, so farm and time performance were not evaluated. Second, quantitative identity-continuity analysis was confined to a 90-min Farm 1 view and conditioned on inherited detections. The audit did not estimate Stage D false-merge rate, identity purity, or cross-camera identity, and the denser Farm 2 and Farm 3 reviews were qualitative. Third, the downstream comparison lacked a complete cow-level crosswalk and aligned graph arrays. It therefore assessed sensitivity of trajectories, marginal summaries, and selected regional evidence, not support, weighted, or mesoscopic graph fidelity. Raw video remains restricted, and the current public code tree is a partial development release rather than a complete reproduction package.

These boundaries narrow the claims but strengthen the contribution. DairySocialTwin does not present a verified map of enduring cow relationships. It offers a general architecture for determining where confidence is earned, where uncertainty persists, and how a relational result can be interrogated. The next decisive step is prospective evaluation on an untouched farm with calibrated interaction outputs, independently adjudicated identity labels, false-merge-sensitive metrics, a complete node crosswalk, aligned adjacency arrays, and paper-matched archival code.

Those additions would extend the present audit from representational readability and qualitative sensitivity to quantitative support, weighted, and mesoscopic fidelity.

## Conclusion

Multi-agent video should be judged where inference terminates, not only where prediction begins. DairySocialTwin follows the evidence chain from a frozen clip ledger and inherited trajectories through interaction modeling, camera-sequence identity reconciliation, observation opportunity, graph summaries, and identity-level video trace-back. Reconciliation improved continuity across all tested gap definitions, yet detector contamination and assignment-sensitive summaries persisted. Cleaner trajectories can make a network interpretable without establishing cow-level relational fidelity. By matching each claim to its evidential tier, the framework turns uncertainty from a hidden liability into an auditable property of video-derived social graphs.

## Experimental Procedures

### Study design and analytical scope

This study analyzed previously recorded video from three commercial dairy farms in Atlantic Canada. Detection boxes, initial tracker identifiers, and 27-point pose trajectories were inherited from an earlier pose-based pipeline. The present analysis contributed clip curation, comparison of interaction models, four-stage camera-sequence identity reconciliation, camera-specific semantic maps, opportunity-conditioned graph construction, camera-local descriptors, manual identity review, matched downstream sensitivity analysis, and an evidence dashboard. The inherited detector and pose estimator were neither retrained nor benchmarked. All conclusions are conditional on their outputs.

The analytical unit changed across evidence tiers. Source clips were the unit for the frozen development split; stabilized segments were the unit for model confusion matrices; tracklets and manually corrected trajectories were the units for identity-continuity evaluation; and camera-local identity pairs within matched observation windows were the units for graph construction. These units were recorded separately to prevent source clips, generated segments, tracklets, and graph observations from being treated as interchangeable sample counts.

### Video chronology, clip curation, and frozen index

The source corpus contained 203 videos totaling 122 h 23 min and 2.05 TB, recorded through five fixed cameras at each farm. Default GoPro filenames and operating-system creation times did not reliably preserve order. Relative chronology was reconstructed from embedded TCD/tmcd 29.97-fps drop-frame timecode and manually checked against sequence relationships. Candidate generation produced 839 clips. Manual review removed 264 clips without usable cattle or motion, 116 during quality control, and 12 during trajectory stabilization. The retained archive contained 447 clips. Forty-seven retained clips were not included when the experimental index was frozen, leaving 400 source clips: 57 friendly, 107 unfriendly, and 236 no interaction. The frozen split contained 330 training source clips and 70 development-validation source clips. Stabilization generated 75 validation segments from the 70 validation sources. Overlapping segments from the same source were not treated as independent source clips.

Seventy-one previously labelled legacy clips were included within the 400-source index. Their interaction labels and trajectories passed the same region annotation, quality-control, and stabilization procedures, but reliable farm, camera, and original timecode metadata were unavailable. Missing fields were explicitly coded as unavailable. These clips contributed to overall development but not to farm- or time-stratified evaluation. The study therefore reports no performance estimate by farm or time of day.

Operational labels were friendly, unfriendly, and no interaction. Friendly and unfriendly referred to the project's visual interaction categories, not inferred affect or welfare state. Annotation records did not preserve an independent second-annotator agreement estimate or a complete parent-encounter grouping audit; these absences define the scope of the development results.

**Interaction-model development**

Five alternatives were compared: a direct three-class trajectory classifier, a cascaded interaction gate followed by friendly-versus-unfriendly classification, a non-graph Transformer, a temporal graph neural network, and an NRI-style latent-edge model.[63,64] Every model was evaluated against the same 75 stabilized validation segments. Accuracy, macro-F1, balanced accuracy, weighted-F1, and class-specific precision, recall, and F1 were recalculated from the reported integer confusion matrices.

The cascaded model had the strongest aggregate development metrics. The NRI-style model was retained for downstream analysis because it recovered more friendly validation segments and produced dyadic latent-edge scores compatible with the graph estimand. This was an exploratory model-selection decision. No independent test set, repeated-seed distribution, or preregistered selection rule was available.

For each eligible dyad and processed time point, the retained model produced friendly score $p_{r,t}$ and unfriendly score $p_{u,t}$. The scores were not calibrated. They were therefore preserved as relative model evidence and were not interpreted as event probabilities or behavioral seconds.

**Camera-specific semantic maps and geometric eligibility**

One representative frame from each camera view was annotated with accessible functional regions, barriers, and circulation paths. Region labels included water, food, rest, path, and water-adjacent congregation where visible. Maps were defined in image coordinates because camera intrinsics, lens-distortion coefficients, surveyed landmarks, and a common metric floor plan were unavailable.

Bounding-box centers were assigned to semantic polygons. When a reference rectangle was used, points were mapped through a bilinear quadrilateral transform. A candidate dyad was eligible only when both assigned identities were visible and the segment joining their mapped centers did not cross an annotated barrier. Eligibility described camera geometry, not physical contact.

**Four-stage camera-sequence identity reconciliation**

Stage A formed conservative micro-tracklets from consecutive frames. Exact repeated boxes were linked directly. Otherwise, recent box motion was estimated, and candidate links were screened by normalized center residual and area ratio. The center-residual and area-ratio thresholds were 0.75 and 1.8. The design favored false splits that later stages could reconnect because an early false merge is difficult to reverse.

Stage B represented each micro-tracklet by an appearance prototype. DINOv2 and MegaDescriptor-L-384 were compared using self-supervised positive pairs from reliable within-tracklet observations and simultaneous-box negatives.[65,66] MegaDescriptor-L-384 was retained. Crop-level embeddings were L2-normalized and combined through quality weights reflecting overlap with other detections, receptive-field gain, relative crop size, image variation, and distance from the frame boundary.

Stage C trained a lightweight short- and long-gap link model without permanent identity labels. Splitting reliable micro-tracklets supplied pseudo-positive pairs; concurrent distinct tracklets supplied pseudo-negative pairs. Candidate links had to progress forward in time, avoid temporal overlap, and span no more than 5 s. Accepted links were manually checked in the Farm 1 audit.

Stage D assigned the remaining tracklets to anonymous camera-sequence identity slots through constrained weighted bipartite matching. Dense Hungarian assignment was used for smaller problems, and a sparse HiGHS formulation was used when scale made dense assignment impractical.[67,68] Temporal overlap was prohibited. Capacity was set to 62 identities for Farm 1, 217 for Farm 2, and 168 for Farm 3. These values combined confirmed herd-size priors with slack for residual non-cow or multi-cow detections. The output count was therefore prior-conditioned and was not used as an accuracy metric. Identities were anonymous and local to ordered camera sequences; no cross-camera or permanent-animal identity was claimed.

**Manual identity reference and continuity metrics**

A review interface displayed proposed identity occurrences and allowed the reviewer to accept an assignment, correct the anonymous identity, or mark a detection invalid. Four clips covering Farm 1, Camera 1; Farm 2, Camera 5; and Farm 3, Camera 1 were reviewed. Farm 1 tracklets were reviewed individually. The denser Farm 2 and Farm 3 clips were reviewed in automatically proposed groups while preserving the underlying tracklet counts. Review was not blinded to algorithmic proposals.

Quantitative continuity was evaluated in the 1:30:44 Farm 1, Camera 1 reference. For a gap threshold $g$, consecutive valid observations of one manually corrected trajectory belonged to the same evaluation block when their effective frame interval did not exceed $g$. Frames in which the active reference trajectory was marked invalid were neutral and did not create a gap. If an evaluation block contained $m$ assigned identity segments, it contributed $m-1$ switches and $m-1$ fragmentation events. Switches were normalized by reviewed hours, fragmentation was averaged over manually corrected trajectories, and continuous duration was the elapsed length of uninterrupted valid runs carrying one assigned identity. Metrics were calculated at $g \in \{1,15,30\}$ frames.

Stage C accepted-link correctness was the fraction of accepted links whose two tracklets belonged to the same manually corrected trajectory. This metric did not evaluate unaccepted links or false merges introduced by Stage D.

**Score-weighted evidence and observation opportunity**
For an anonymous identity pair $(i, j)$ at processed time $t$, let $c_t$ be the product of both inherited bounding-box confidences, $\delta_t$ the elapsed source-video time represented by that observation, and $\Delta_t$ an eligibility indicator equal to 1 when both identities were visible and satisfied the camera-specific geometric rules and 0 otherwise. Friendly evidence $e_r$, unfriendly evidence $e_u$, and confidence-weighted opportunity $o$ were accumulated as

$$e_r(i,j) = \sum_t \Delta_t\, p_{r,t} c_t \delta_t, \qquad e_u(i,j) = \sum_t \Delta_t\, p_{u,t} c_t \delta_t,$$

$$o(i,j) = \sum_t \Delta_t\, c_t \delta_t.$$

When $o(i,j) > 0$, opportunity-normalized evidence was

$$n_r(i,j) = \frac{e_r(i,j)}{o(i,j)}, \qquad n_u(i,j) = \frac{e_u(i,j)}{o(i,j)}.$$

Pairs with $o(i,j) = 0$ were retained as unobserved, not assigned zero evidence. Friendly and unfriendly layers remained separate and non-negative. Negative plotting direction in some figures was a visual convention only.

**Camera-local descriptors**
The observed dwell-time budget recorded the total visible time assigned to each functional region. For identity $i$, camera-local alone fraction $\alpha_i$ was the proportion of its visible time for which it was the sole observed cow in its assigned region. "Alone" therefore referred to camera visibility, not absence of social behavior.

Friendly strength was the sum of $e_r(i,j)$ over observed partners. After normalization by the identity's visible time, the low-friendly-strength index was

$$\beta_i = 1 - \frac{N_r(i)}{\max_{x \in C} N_r(x)}.$$

If every identity had zero normalized friendly strength, $\beta_i$ was set to 1 for all identities to avoid division by zero and to preserve the meaning of uniformly low evidence. Partner-weight evenness $\gamma_i$ was normalized Shannon entropy over the identity's nonzero friendly partner weights.[69] If the number of nonzero partners was 0 or 1, $\gamma_i$ was set to 0. The exploratory isolation index was

$$I_i = 0.4\alpha_i + 0.3\beta_i + 0.3(1 - \gamma_i).$$

Each component was retained alongside the composite. Region-weight evenness was calculated separately for friendly and unfriendly evidence. Community summaries used Louvain partitioning of the friendly evidence graph within camera-local time windows.[70] Community labels were not compared directly across windows; pairwise co-membership was used because numerical community labels are arbitrary. These descriptors were exploratory and were not validated as welfare or personality measures.

**Matched downstream sensitivity analysis**
For every reviewed record, downstream summaries were regenerated under manually corrected and reconciled identity assignments while the observation window, semantic map, inherited detections, and interaction scores were fixed. Comparisons covered trajectories, visible-time budgets, isolation components, and aggregate opportunity-conditioned evidence by region. Raw tracker labels supplied a fragmentation reference.

The comparison was qualitative. Panels used independently ordered anonymous identities, and neither a complete cow-level crosswalk nor aligned numerical adjacency arrays were retained. Node support, edge precision or recall, weighted-edge error, strongest-partner retention, node-strength rank correlation, and community agreement were therefore not estimated. No raster-derived support statistic was used as a substitute.

**Evidence dashboard and data handling**
The dashboard rendered stored numerical summaries for user-selected anonymous identities. Selecting an identity-level point opened recorded appearance intervals; selecting an interval opened the corresponding farm, camera, clip, and frame in the review interface. Herd-wide figures were not recomputed by identity filtering. The trace-back path connected selected identity summaries to observational context, not every edge to every contributing frame.

Analytical rows retained farm-camera sequence, anonymous local identity, clip, frame or appearance time, and the quantities used by the dashboard. Source-video hashes and checkpoint-version links were not displayed in the provenance card, so the dashboard should be understood as observational trace-back rather than complete computational lineage.

**Statistical analysis**
Classification metrics were recalculated from integer confusion counts for 75 stabilized validation segments. Identity-continuity changes were calculated relative to the manually corrected Farm 1 reference at all three reported gap thresholds. Stage C accepted-link correctness was reported as an exact numerator and denominator. Downstream manual-reconciled comparisons were descriptive because aligned identity and graph arrays were unavailable. No null-hypothesis tests were performed. No confidence interval is presented where repeated seeds, cluster-aware resampling units, or independent annotations were unavailable. The analysis emphasizes exact counts, transparent denominators, and claim-matched interpretation.

## Resource Availability

**Lead contact**
Further information and requests should be directed to the corresponding author, Suresh Raja Neethirajan (Sureshraja@dal.ca).
**Materials availability**
This study did not generate new biological reagents.

**Data and code availability**
Public code is available in the DairySocialTwin repository at

https://github.com/mooanalytica/DairySocialTwin under the MIT License. The repository includes author-generated tools for CSV segmentation, directed-interaction and interaction annotation, quality control, trajectory stabilization, camera-specific semantic mapping, NRI-style and non-graph temporal modeling, camera-sequence re-identification, and the dashboard and social-network interface. The immutable public code state reviewed for this manuscript is commit 4432f31f63eff53698addb45d65b3e136f07d03d.

The 447-clip curated archive, 400-clip frozen experimental index, manually corrected identity-review clips, trained checkpoints, full configurations, processed results, and highest-resolution figure sources are retained in controlled institutional storage. Raw farm video and site-identifying metadata cannot be publicly redistributed. Eligible de-identified trajectories, labels, manifests, identity-review outputs, graph tables, and figure-source data can be provided to editors and reviewers subject to participating-farm agreements. Requests should be directed to the lead contact and will be assessed under the governing agreements and institutional requirements. Because the controlled assets are not distributed with the public code, the repository supports inspection of the author-generated workflow but is not, by itself, a complete reproduction package.

**Acknowledgments**

The authors thank the participating commercial dairy farms for access to operational video and contextual discussion, and members of the MooAnalytica laboratory for infrastructure and project support. This work was supported by the Natural Sciences and Engineering Research Council of Canada.

**Author Contributions**

Y.H.: Methodology, Software, Validation, Formal Analysis, Investigation, Data Curation, Visualization, and Writing - Original Draft. S.R.: Conceptualization, Supervision, Resources, Project Administration, and Writing - Review and Editing.

**Declaration of Interests**

The authors declare no competing interests.

**Ethics and Permissions**

The study involved secondary analysis of previously recorded, non-interventional video from commercial dairy farms. The original data-collection protocol was approved by the Dalhousie University Animal Care and Use Committee (Protocol 2024-026; approved 16 May 2024) and complied with Canadian Council on Animal Care guidelines. Data acquisition was entirely non-invasive and limited to passive image and video recording, without physical contact, experimental treatment, or alteration of routine farm management. Participating farm owners provided written informed consent after receiving a full explanation of the study objectives and procedures. Video access, storage, and use were governed by farm agreements. Raw footage and site-identifying information remain restricted, and manuscript figures were selected or cropped to exclude identifiable farm personnel. The present secondary analysis introduced no additional animal procedures.